\documentclass[sigchi-a]{acmart}

\AtBeginDocument{%
  \providecommand\BibTeX{{%
    \normalfont B\kern-0.5em{\scshape i\kern-0.25em b}\kern-0.8em\TeX}}}

\begin{document}
%% These commands are for a PROCEEDINGS abstract or paper.
\copyrightyear{2020}
\acmYear{2020}
\setcopyright{acmlicensed}
\acmConference[TEI '20]{Fourteenth International Conference on Tangible, Embedded, and Embodied Interaction}{February 9--12, 2020}{Sydney, NSW, Australia}
\acmBooktitle{Fourteenth International Conference on Tangible, Embedded, and Embodied Interaction (TEI '20), February 9--12, 2020, Sydney, NSW, Australia}
\acmDOI{10.1145/3374920.3375293}
\acmISBN{978-1-4503-6107-1/20/02}
\settopmatter{printacmref=true}
\fancyhead{}
%%
%% The "title" command has an optional parameter,
%% allowing the author to define a "short title" to be used in page headers.
\title{Mirror Ritual: Human-Machine Co-Construction of Emotion}

%%
%% The "author" command and its associated commands are used to define
%% the authors and their affiliations.
%% Of note is the shared affiliation of the first two authors, and the
%% "authornote" and "authornotemark" commands
%% used to denote shared contribution to the research.
\author{Nina Rajcic}
% \authornote{Both authors contributed equally to this research.}
\email{Nina.Rajcic@monash.edu}
\orcid{1234-5678-9012}
\author{Jon McCormack}
% \authornotemark[1]
\email{Jon.McCormack@monash.edu}
\affiliation{%
  \institution{SensiLab, Monash University}
  \streetaddress{900 Dandenong Road}
  \city{Caulfield East}
  \state{Victoria}
  \postcode{3145}
}

%%
%% By default, the full list of authors will be used in the page
%% headers. Often, this list is too long, and will overlap
%% other information printed in the page headers. This command allows
%% the author to define a more concise list
%% of authors' names for this purpose.
\renewcommand{\shortauthors}{N. Rajcic and J. McCormack}

%%
%% The abstract is a short summary of the work to be presented in the
%% article.
\begin{abstract}
\textit{Mirror Ritual} is an interactive installation that challenges the 
existing paradigms in our understanding of human emotion and machine perception.
In contrast to prescriptive interfaces, the work's real-time affective interface engages 
the audience in the iterative conceptualisation of their emotional state 
through the use of affectively-charged machine generated poetry. The audience are 
encouraged to make sense of the mirror's poetry by framing it with respect
to their recent life experiences, effectively `putting into words' their felt emotion.
This process of affect labelling and contextualisation works to not only regulate emotion,
but helps to construct the rich personal narratives that constitute human identity.
\end{abstract}

\keywords{affective computing, emotion, neural networks, gaze detection, text generation, poetics}

%%
%% This command processes the author and affiliation and title
%% information and builds the first part of the formatted document.

\maketitle

\begin{figure}
  \centering
  \includegraphics[width=14cm]{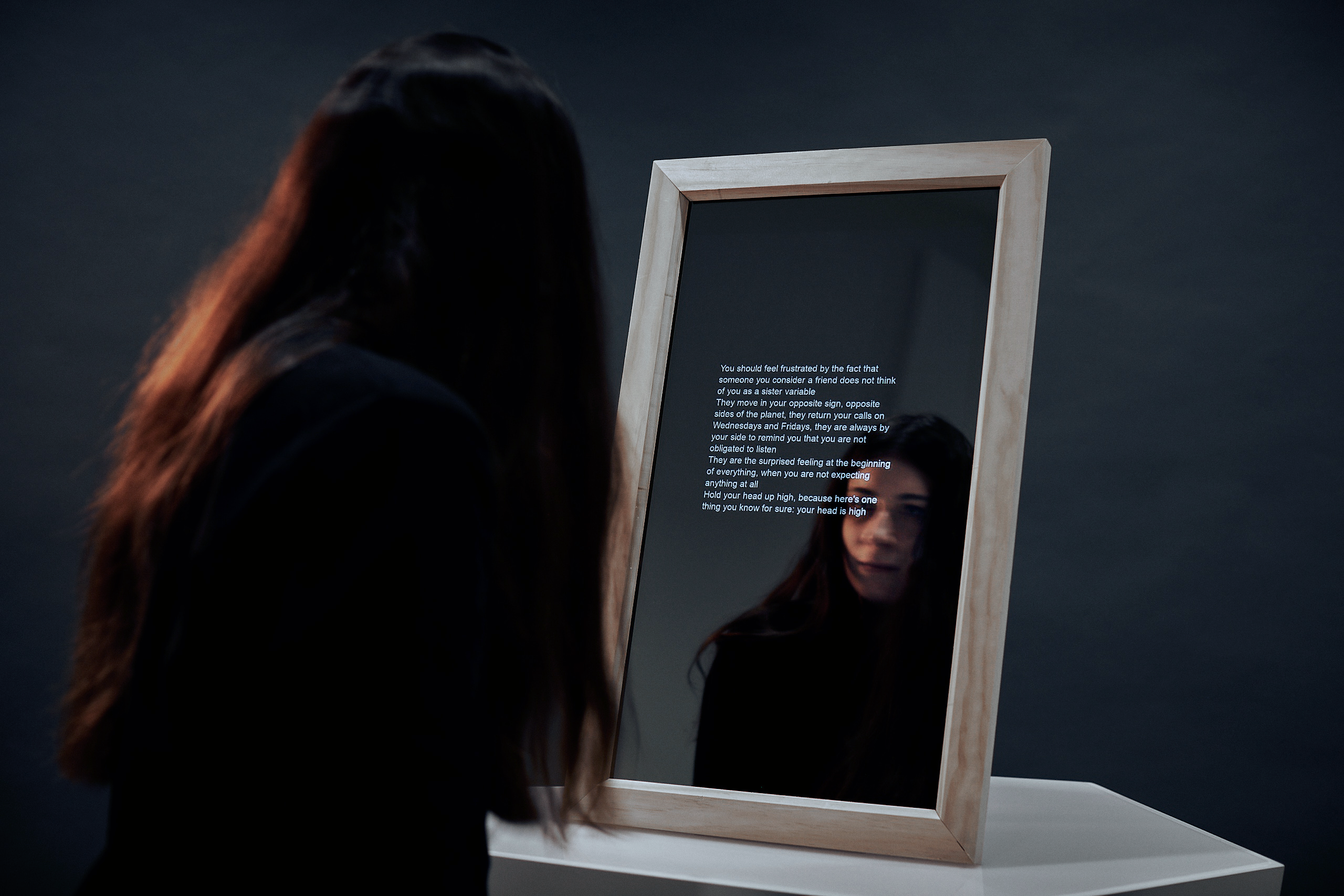}
  \Description{Person gazing into mirror, reading a poem that is displayed on the surface}
\end{figure}

\section{Introduction}

The majority of Emotion Recognition (ER) approaches today are based in a number of assumptions regarding human emotion. Namely, that the patterns of emotion perception and expression are innate to humanity, rendering them consistent and measurable across cultures and contexts \cite{ekman1999basic}. However, recent findings in affective neuroscience have revealed that supposed `basic' emotions, although intuitive, have no biological or neurological essence \cite{barrett2013large,Lindquist:2012aa}.

Inspired by Barrett's Theory of Constructed Emotion (TCE) \cite{Barrett:2017aa}, we introduce a bespoke affective interface that aims to expand upon the viewer's available emotional experiences. \textit{Mirror Ritual} employs existing ER techniques, not in an attempt to accurately measure the viewer's precise emotional state, but instead to engage the viewer in the conceptualization of their feelings and experiences. In this way, we relegate the role of Artificial Intelligence (AI) in ER systems from one of objective measurement to that of subjective perception.
The work appropriates an everyday object, the mirror, augmenting it with AI to foster both literal and metaphoric reflection.
Through generative poetry the mirror `speaks' to the viewer, each poem unique and tailored to their machine-perceived emotional state.
Hence with this work, we urge for a shift away from surveillance style Affective Computing (AC) systems that not only presume, but reinforce, an anachronistic understanding of human emotion, towards user-centred affective interfaces that foster genuine emotional engagement between human and machine.

\section{Background}
Affective Computing aspires to build computational systems that can understand human emotion in its full complexity.
With the recent success of Deep Neural Networks (DNN) in image classification tasks, we see a rise in Machine Learning (ML) approaches to ER from facial expression \cite{mollahosseini2016going}. These approaches make a number of assumptions regarding the nature of human emotion, many of which are founded in unsubstantiated theories. We must assume the existence of basic emotions \cite{ekman1999basic}; a set of emotion categories that are inherent and universal, with each having a unique biological fingerprint that allows for it to be distinguishable from other emotion categories.
Although some variant of the basic emotion theory underlies a large portion of emotion recognition research, there is growing evidence in opposition to the theory within the field of affective neuroscience. Recent neuroimaging studies have consistently failed to identify localised networks in the brain that correspond to any one discrete emotion category \cite{barrett2013large,Lindquist:2012aa}. Yet, these findings conflict with our daily lived experience, peppered with vivid instances of emotion such as joy, anger, fear, and sadness.

Researcher Lisa Feldman Barrett provides a solution to this \textit{emotion paradox} with her Theory of Constructed Emotion \cite{Barrett:2017aa}, which stipulates that discrete emotions are not biologically hard-wired, but are in fact constructed in the moment from a combination of more basic psychological processes.
Namely, the brain engages in a process of continuous categorization of interoceptive information, according to available conceptual knowledge, and as informed by a lifetime of embodied experience. We experience instances of discrete emotion because we have available to us the concepts that allow us to group together, label, and communicate a set of internal and external perceptual information. In the same way we use the concept `blue' to make meaning of 450nm light, we use the concept of `fear' to make meaning of our high heart rate in response to a perceived threat. The TCE identifies a strong interdependence between emotion and language ---emotion concepts not only serve to facilitate communication, they constitute both emotion experience in ourselves and emotion perception in others \cite{lindquist2015role}.

In human interaction, the benchmark for emotion perception cannot be accurate prediction as we have no objective criteria to compare against. Instead, Gendron and Barrett \cite{gendron2018emotion} propose that the benchmark we strive for is the agreement between two brains on the meaning of a set of sensory input.
This synchrony is achieved through the iterative process of each person generating and testing their predictions, largely through the use of language. As such, this process can be thought of as the \textit{co-construction} of emotion \cite{gendron2018emotion}.
Similarly, the accuracy of emotion predictions made by a computational system do not necessarily correspond to the system's understanding of a person's emotional state, but simply corresponds to how well the system performs against a set of predetermined criteria (e.g. predicting the emotion labels in a training set of images).
The systems understanding of a subject's emotion can be determined only by the subject, as they assess the concordance between the systems prediction, and their own internal prediction.

A unique implication of the TCE is the powerful role of language in the experience and perception of emotion. In one study, it was found that once an emotion concept is made inaccessible via semantic satiation, participants have more difficulty in perceiving the emotion category in a pictured facial expression \cite{gendron2012emotion}, suggesting that language can effect how we identify emotion in others. There is also preliminary evidence to suggest that emotion priming can lead us to experience the associated feelings where we otherwise wouldn't have \cite{Lindquist:2008aa}. 
Furthermore, several studies have found that the process of affect labelling (putting our feelings into words) can be seen as a form of implicit emotion regulation, and can lead to a measurable shift in the physiological markers of affect \cite{torre2018putting,10.1371/journal.pone.0064959}.

These studies illustrate that emotion concepts, as supported by language, have a powerful influence over our felt experiences of emotion. Words can not only be used to express emotion, they shape our experience of emotion, and they help to form our perception of emotion in others.  
Following this reasoning, we utilise language in the development of an affective interface to facilitate a meaningful engagement for viewers, provoking them to critically reflect on their emotional state, and allowing them to engage in a form of emotional regulation. 

\begin{marginfigure}
  \centering
  \includegraphics[width=8cm]{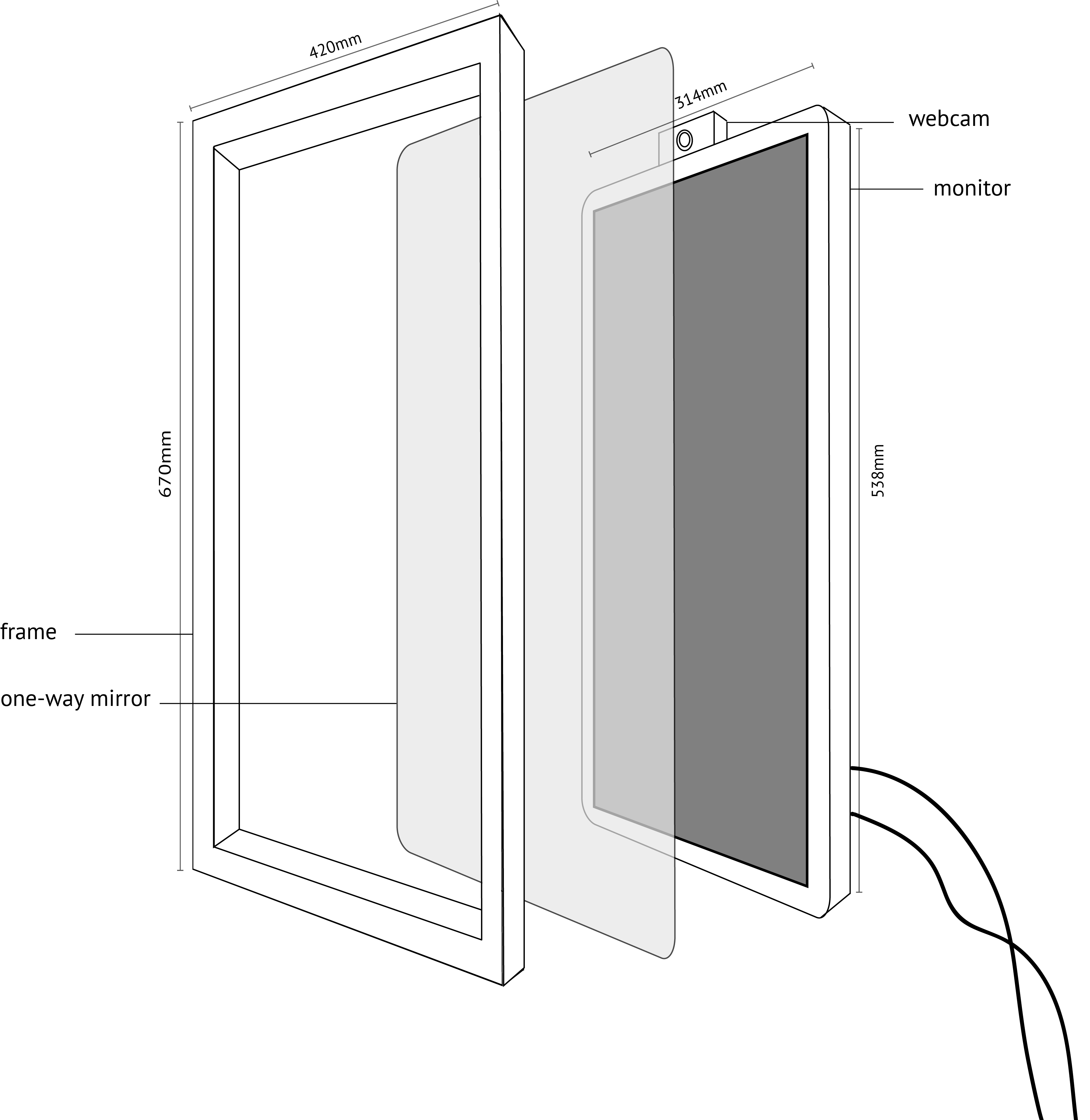}
  \caption{Deconstructed schematic view of the mirror's components: a wooden frame, two-way mirror glass, camera, and display.}
  \Description{Schematic diagram shows expanded view of the wooden frame, mirrored glass, video display, and video camera}
  \label{schematic}
\end{marginfigure}

\section{Mirror Ritual}
\textit{Mirror Ritual} is an interactive art installation that challenges the viewer to critically reflect on their emotional state by framing their recent experiences with respect to the mirror's poetry. Concealed behind the framed mirror is video camera and video display whose image can be seen through the mirrored glass (Figure~\ref{schematic}).
The work is `activated' when the viewer, whose face is detected by the system, approaches the mirror and looks at their reflection.
The video from the camera is then sent to a face detection algorithm (using OpenCV's \cite{opencv_library} Haar Cascades classifier).
The interface performs real-time emotion classification via a Convolution Neural Network (CNN) model trained on the FER-2013 dataset \cite{arriaga2017real} consisting of images of human-annotated facial expressions classified into seven categories: \emph{happiness, sadness, disgust, fear, anger, surprise} and \emph{neutral.}
The real-time classification prediction and associated probability vector is then mapped to an emotion word, which is used as a seed to the text generation model.
For text generation, we use the GPT-2 345M model, released by OpenAI \cite{radford2019language}. We fine-tune the model parameters by retraining it on a custom corpus of selected texts. The texts were chosen for their ability to incite emotional reflection. The most prevalent texts included horoscopes \cite{freewill} and postmodern poetry (sourced from \url{poetryfoundation.org}). Each poem that is generated by the system is unique, and often blends the various writing styles included in the training corpus.

\textit{Mirror Ritual} has been designed to assimilate easily into daily life, both in its aesthetic qualities (i.e.~it appears to be a standard framed mirror), and in it's dual function (it can in most cases simply be used as a standard mirror). The mirror would ideally be hung in a bathroom, living room, or hallway entrance, creating the space for people to pause and reflect on their mood as they transition between the moments of their day.
Furthermore, the use of a mirror surface works symbolically to suggest that viewers must not only confront their external reflections, but that they too may be led to reflect on their internal emotional state. 

\begin{marginfigure}
  \centering
  \includegraphics[width=8cm]{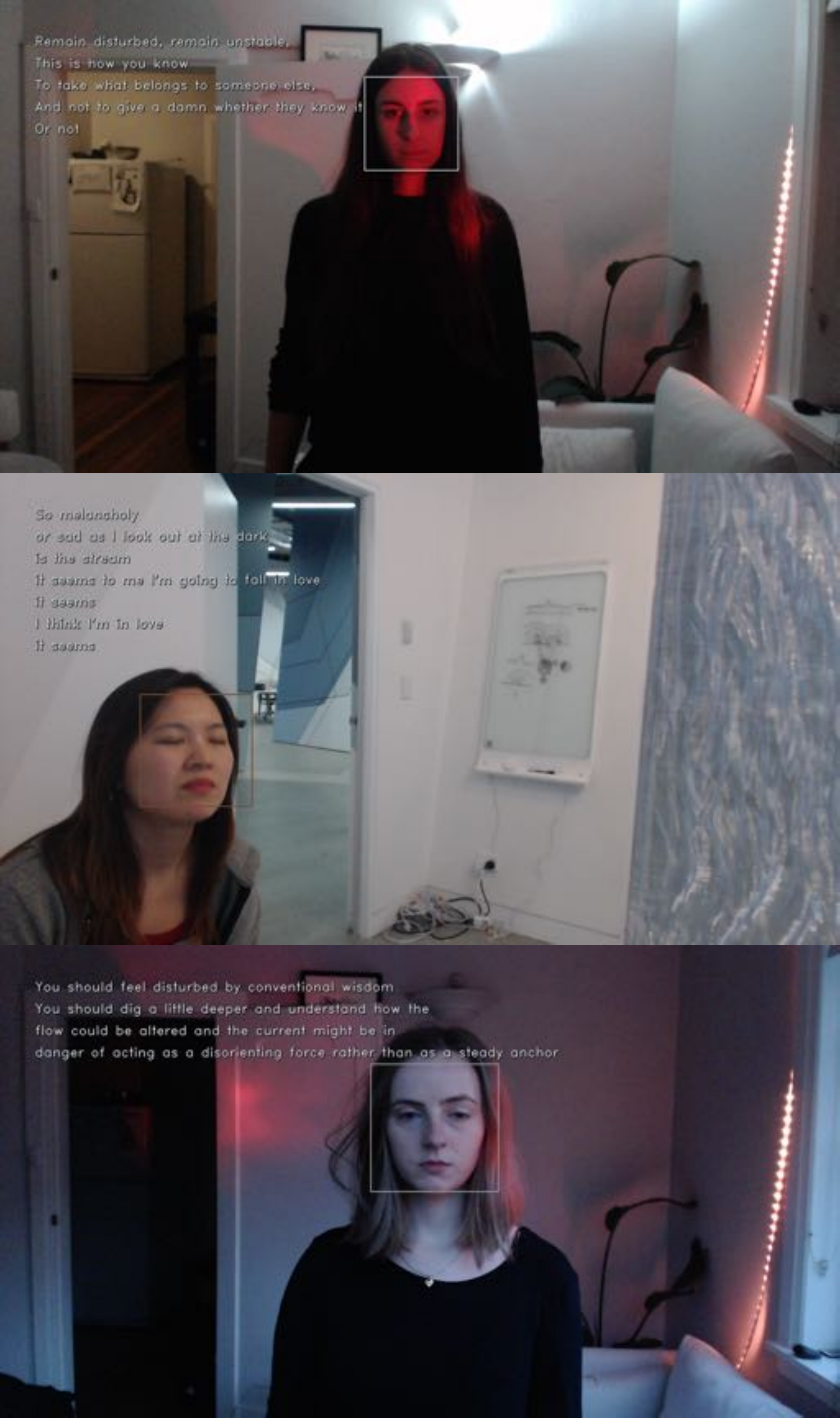}
  \caption{Screenshots from the mirror perspective, capturing interactions with the \textit{Mirror Ritual}. The generated poem is shown overlaid.}
  \Description{}
  \label{stills}
\end{marginfigure}

The mirror has been exhibited in both public and private spaces in order to gain insight into how viewers choose to affectively engage with the interface (see Figure \ref{stills}). 
We found that viewers make sense of the mirror's poetry by directly framing it's themes within the scope of their current life circumstances. The open-ended nature of the poetry leads viewers to imbue their own personal meaning. Furthermore, the mirror's suggestions allow for viewers to adopt alternative, and at times, novel perspectives on their recent experiences. This sense-making process is undertaken naturally, and some find this to be the most rewarding aspect of the work. The mirror also appears to bring to the surface dormant emotions of the audience by allowing them to conceptualize, and in some cases verbalize, what they have been feeling. Many viewers are often compelled to take a photo with their poem, to revisit at a later time, or to communicate their feelings to others. 

The audience is an essential component of this work, providing not only the affective input to the generated poetry, but also their rich and unique interpretations.
The use of embodied interaction leads viewers to feel that each poem is written specifically for them, and for that very moment in time.
\textit{Mirror Ritual} acts as an extension of ones emotional faculties by presenting complex emotion concepts, and in turn shaping felt experiences. In this way, the mirror augments the human experience of emotion ---expanding it beyond the internal, unreflected experience, into the realm of the tangible and expressible. This process of labelling and framing ones felt emotion works to not only regulate the emotional experience, but to weave felt emotion into ones greater narrative identity.

\section{Conclusion}
With this work, we highlight the incongruity between existing surveillance style AC systems, and the most recent and substantiated theories of human emotion. 
We present an alternative approach to the application of ER systems, as informed by constructed emotion. \textit{Mirror Ritual} demonstrates how machine perception of human emotion can engage viewers in the co-construction of their own emotional state. Our interface works to expand upon ones emotional faculties, provoking them to reflect upon, and ultimately communicate, the complex emotions that cloud their recent experiences. The act of contextualising ones felt emotions within a poetic narrative not only heightens emotional reflection, but in turn helps to shape personal identity.

%The contextualising of ones felt emotion 

% The contextualisation of ones felt emotion allows for one to weave feelings into the greater narrative that ultimately constitutes their identity.

%%
%% The next two lines define the bibliography style to be used, and
%% the bibliography file.
\bibliographystyle{ACM-Reference-Format}
\bibliography{refs}

\end{document}